\documentclass[a4paper]{jpconf}
\usepackage{graphicx} % Required for inserting images
\usepackage{amsmath}
\usepackage{amsmath} % for \text{} and enhanced math support
\usepackage{amssymb} % for \lesssim

\usepackage{indentfirst}

\usepackage{amsmath}
\usepackage{eso-pic}
\usepackage{lineno}
\usepackage{tabto}
\usepackage{tikz}
\newcommand{\comment}[1]{}

\begin{document}

\comment{
    \AddToShipoutPictureBG{
    \begin{tikzpicture}[remember picture, overlay,
                        help lines/.append style={
                            line width=0.05pt, color=blue!10},
                        major divisions/.style={
                            help lines,line width=0.1pt, color=red!20} 
                        ]
        \draw[help lines](current page.south west) grid[step=10pt]
                        (current page.north east);
        \draw[major divisions](current page.south west) grid[step=50pt]
                        (current page.north east);
    \end{tikzpicture}
    }
}
\title{Study  of a small-scale gamma-ray detection system employing Compton scattering with a monolithic CeBr$_3$ crystal and segmented photodetector array}

%\author{Veronika Asova$^1$,Galin Bistrev$^2$, Venelin Kozhuharov, Simeon Ivanov}
\author{Veronika Asova$^1$, Galin Bistrev$^1$, Simeon Ivanov$^1,^2$}

\address{$^1$Faculty of Physics, Sofia University, 5 J. Bourchier Blvd., 1164 Sofia, BG}
\address{$^2$National Centre of Excellence Mechatronics and Clean Technologies, Sofia University, 1164 Sofia, BG}

\ead{viasova@uni-sofia.bg, gbistrev@uni-sofia.bg}
\date{November 2024}

\begin{abstract}
Study of high energy cosmic events in the MeV range  requires detector with high efficiency and energy resolution to be constructed. The present setup consisting of   scintillator crystals $CeBr_3$ with different thickness, each coupled with 12 x 12 segmented SiPM-based photodetector in a multichannel system represents an initial exploration of a gamma imaging system based on Compton scattering principles. A Monte Carlo simulation was conducted to evaluate the energy deposit and detection efficiency using the  $^{137}$Cs gamma line.
The study reveals a correlation between the relative distance between detector planes and the energy deposition efficiency, providing valuable insights into optimizing the telescope’s design.   
\end{abstract}

\section{Introduction}
Gamma ray measurements are a topic of major importance in several fields of science and technology e.g.  medical imaging, environmental monitoring,
study of high-energy cosmic events, etc~\cite{bib:gamma-astro},~\cite{bib:gamma-env},~\cite{bib:gamma-med}.
 
 The detection of high-energy cosmic events in the MeV gamma-ray range is effectively achieved using Compton telescopes, which offer high detection efficiency and excellent energy resolution. 
 They are preferred due to the  coincidence requirement and coarse imaging ability, 
 which greatly suppress background from cosmic ray interactions in the instrument and spacecraft ~\cite{bib:gamma}.
 
 Compton telescopes work on the principle of Compton scattering, 
 which is   the main process through which gamma quanta  interact with matter in the MeV ranges. 
 Measuring the position and deposited energies of a sequence of such scatters will provide a cone of possible directions for the incoming gamma quanta. 
 By combining data from many detected events, 
 the telescope is able to create an image of the gamma source.
 
 For the construction of such device a detector with high light yield, 
 good stopping power, fast time response is required.  
 Scintillator detectors satisfy these requirements,
 while also offering the benefits of room temperature operation and relatively low costs ~\cite{bib:Compton-scin}.

 Inorganic scintillators like $CeBr_3$ are being evaluated for use in Compton telescopes and nanosatellites due to their high yield ($\sim$68 000 ph/MeV), 
 short light decay time (17 ns) and decent resolution 
 (full width at half maximum over the peak position)  $\sim$ 4\% at 662 keV  ~\cite{bib:CeBr3prop}, ~\cite{bib:pdg}.
 Another promising inorganic scintillator crystal is the $LaBr_3:Ce$, 
 which has high light yield ($\sim$61 000 ph/MeV),
 high energy resolution (\( 2.85~\% \pm 0.5~\% \text{ at 662 keV} \)) 
 and short decay time (26 ns) ~\cite{{bib:La Br_3 Ce}}, 
 but due to the natural radioactivity of lanthanum, 
 it is not useful in low-noise environments~\cite{{bib:La Br_3 low noise}}.
 Even though at the cost of sensitivity  natural radioactivity of $LaBr_3:Ce$ provides on-board energy  calibration of space-borne detectors where deploying external calibration sources is difficult.

 Due to these characteristics it is essential to gain a better understanding of Compton telescope constructed using $CeBr_3$ based scintillator detector.
 In the current article, the 
% sets of measurements  
results
 presented are 
% obtained using 
for
 a Compton-ray telescope utilizing two  $CeBr_3$ crystals, each of them coupled to a SiPM-based photo detector.
 The constructed and the simulated experimental setup is presented in the subsequent section.  
 
\section{Compton imaging}
Reconstructing the trajectory of gamma quanta relies on the usage of  Compton kinematics in order to determine the original direction of a photon that has interacted multiple times across two or more parallel detector planes. 

In most cases two detector planes are used- a front scatter detector plane, where the Compton scattering occurs, and a rear absorber detector plane that captures the scattered gamma ray ~\cite{bib:Compton-plane}.  

For an event, in which a  Compton scattering is observed in the first plane 
followed by a photo absorption in the second one, 
the scattering angle $\theta_C$ is given by the formula ~\cite{bib:Compton}:
\begin{equation}
    \cos \theta_C = 1 + {m_e c^2( {\frac{1}{E_t}-\frac{1}{E_2}}})
\end{equation}
where \[E_t = E_1+E_2\] is the total energy, $E_1$ is the energy  of the Compton scattered gamma quant and $E_2$ is the energy of the absorbed Compton scattered gamma quant. Here the mass energy of the electron is  $m_e c^2$.

Therefore, the emitter's position is reconstructed up to an arbitrary azimuthal angle, meaning it lies on the surface of a cone with an axis along the line between $x_1$ and $x_2$ and the angle $\theta_C$.
By combining multiple such cones, the location or spatial distribution of the gamma-ray source can be accurately determined.

For higher-energy gamma photons, the probability of escaping the second detector without full energy deposition increases. 
In such cases, a three-stage Compton camera may provide a better reconstruction, though at the cost of reduced detection efficiency ~\cite{bib:three_plane}.
 
\section{Experimental setup}
\begin{figure}[!htp]
    \begin{center}
        \includegraphics[width=0.9\textwidth]{
            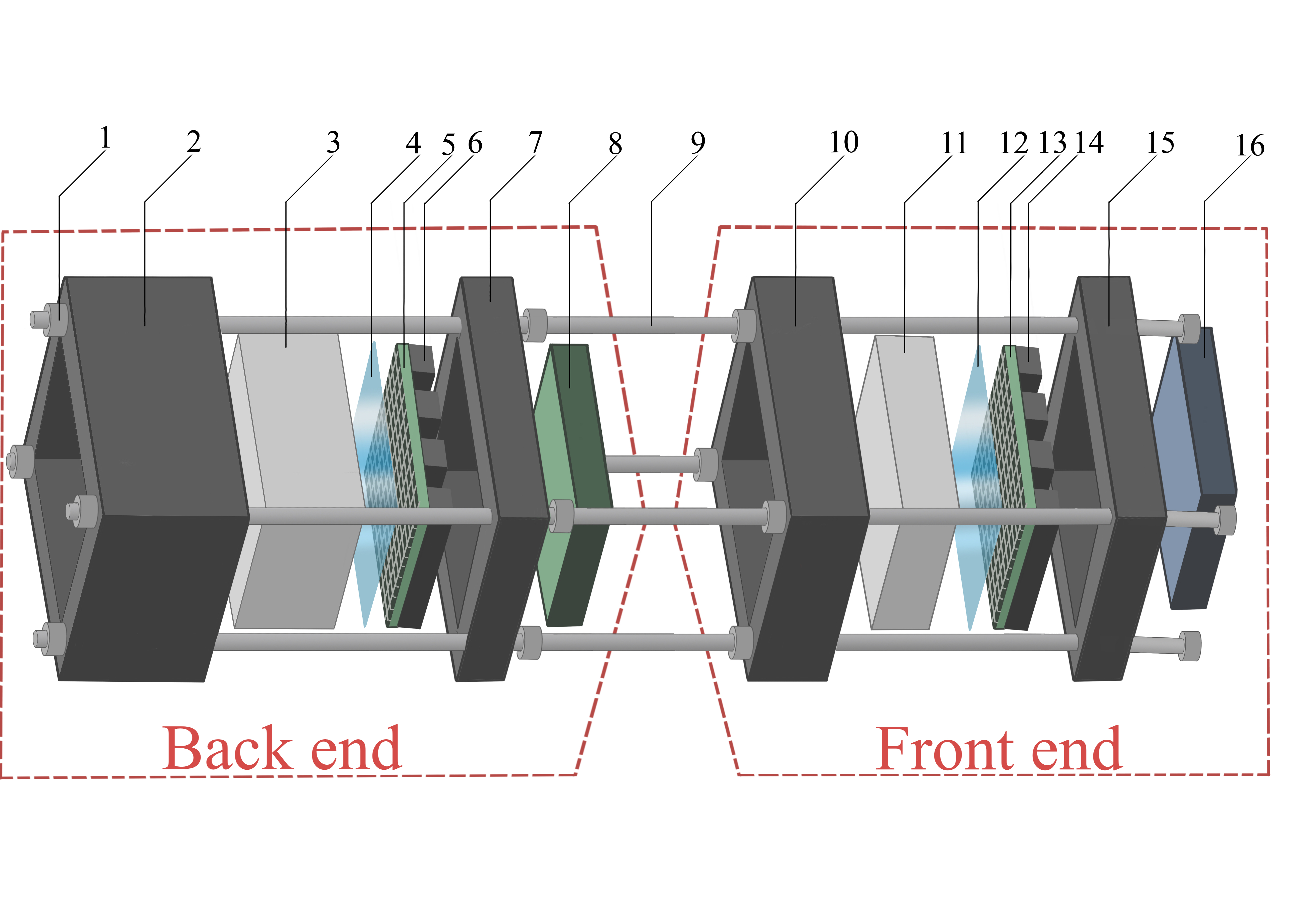
        }
        \caption{
            \label{fig:housing}
            A schematic representation of the experimental setup. 
       }
    \end{center}
    \end{figure}

The current experimental setup consists of two SiPMs, each coupled to a CeBr\textsubscript{3} crystal with different dimensions: 
\( 51 \times 51 \times 25 \, ~\text{mm}^3 \) and \( 51 \times 51 \times 10 \, ~\text{mm}^3 \).
Further in this paper, the former will be referred to as ''thick crystal'' and the latter as ''thin crystal''.
The choice of SiPMs over more traditionally used PMTs due to the former's  combination of high gain, favorable signal-to-noise ratio, and low operating voltage.
The front end of the Compton telescope incorporates the thick crystal,
while the back end incorporates the thin one. 
Two different readout systems are constructed for the two ends of the telescope. 
The Compton telescope and its components are shown on Figure  \ref{fig:housing} where the numbers point to:

\begin{itemize}
        \item{ 1 - Fasteners  }
        \item{ 2 - 3D printed upper casing for thick crystal }
        \item{ 3 -$CeBr_3$ thick scintillator  crystal }
        \item{ 4 - silicon optical grease }
        \item{ 5 - SiPM matrix }
        \item{ 6 - Samtec 80-way connectors type QTE-040-03-F-D-A }
        \item{ 7 - 3D printed lower casing }
        \item{ 8 - AiT AB424T-ARRAY144P Tileable 4+24 Channel Hybrid
               Active Base }
        \item{ 9 - metal rod }
        \item{ 10 - 3D printed upper casing for thin crystal
               }
        \item{ 11 - 2 - $CeBr_3$ thin scintillator crystal}
        
        \item{ 12 - silicone optical grease}
        \item{ 13 - SiPM matrix }
        \item{ 14 - Samtec 80-way connectors type QTE-040-03-F-D-A}      
        \item{ 15 - 3D printed lower casing }
        \item{ 16 - ARRAYC-30035-144P Breakout Board }

    \end{itemize}

\begin{figure}[!htp]
    \begin{center}
        \includegraphics[width=0.6\textwidth]{
            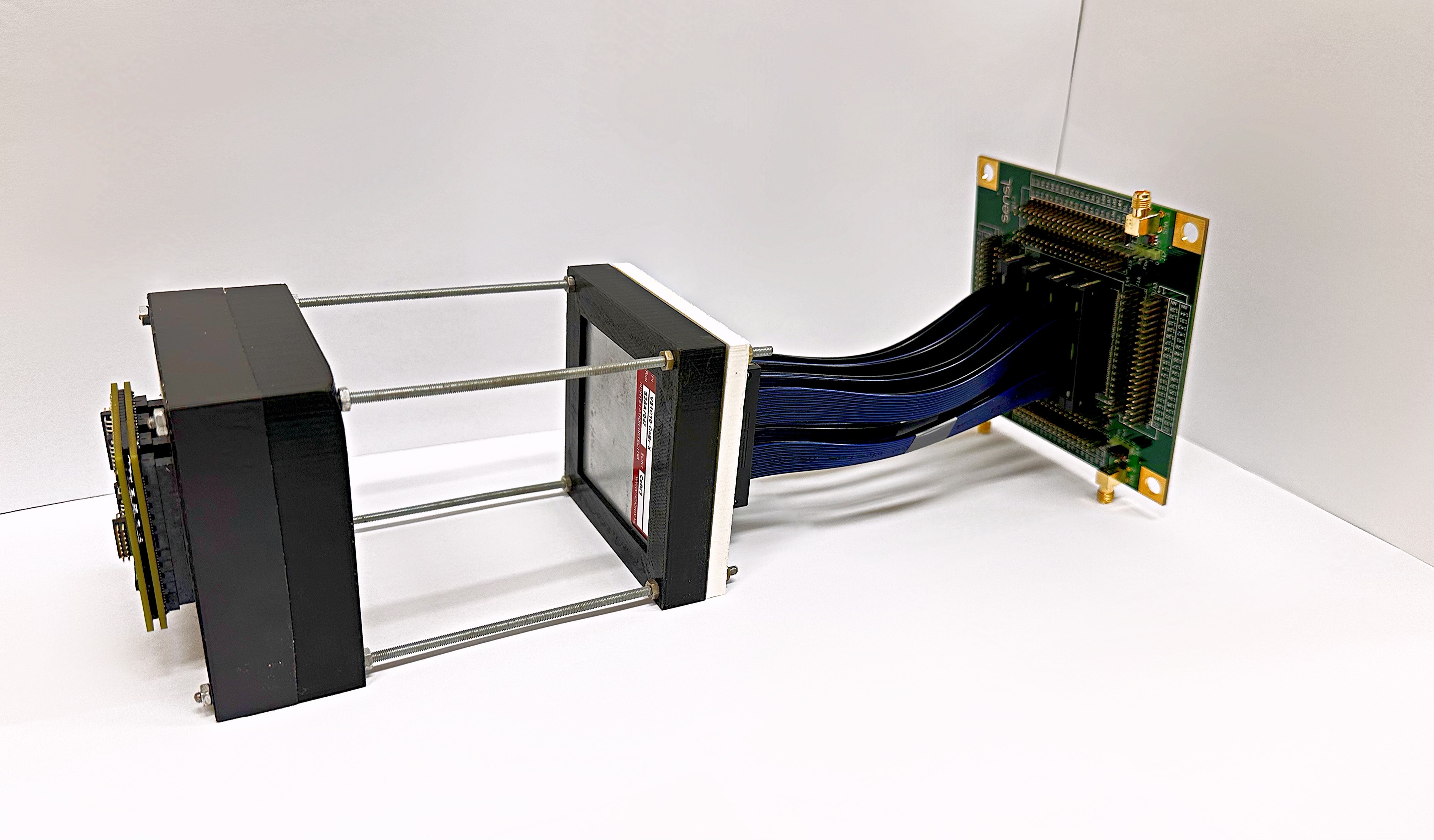
        }
        \caption{
            \label{fig:telescope}
            A picture of the assembled test Compton telescope. 
        }
    \end{center}
    \end{figure}

    \begin{figure}[!htp]
    \begin{center}
        \includegraphics[width=0.6\textwidth]{
            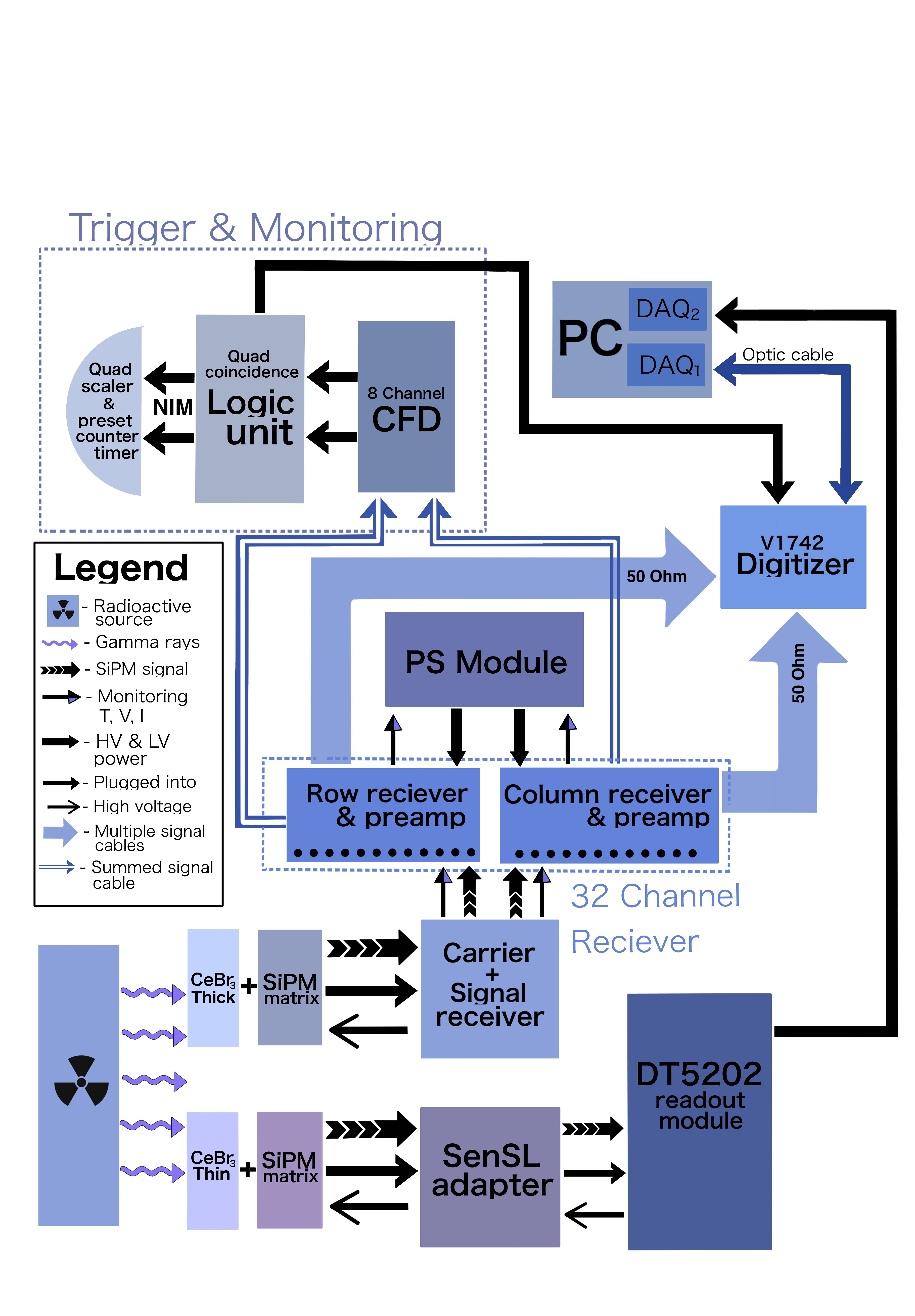
        }
        \caption{
            \label{fig:setup}
            A scheme of the full experimental setup - Compton telescope and the two readout systems responsible for the operation of the two ends of the telescope. 
        }
    \end{center}
    \end{figure}
A real image of the setup is shown on Figure \ref{fig:telescope}. The full setup which incorporated the readout system for both ends is shown on Figure \ref{fig:setup}.
A more detailed description of the front end  can be found at \cite{bib:nafski2022}.

\subsection{Back end}
The thick crystal is coupled to one of the SiPM arrays via a 3D printed housing. 
The whole structure is sealed using a silicone optical grease, 
which reduces reflection losses and improves the overall light transmission between the crystal and the SiPM \cite{bib:optic}.

The ABPS provides both high voltage for the SiPMs and low voltage for powering the frontend electronics and the amplifiers.
The overall power consumption of the back-end electronics is of the order of milliwatts $\mathcal{O}(\mathrm{mW})$, 
which makes it suitable for power-constrained environments such as CubeSats or nanosatellites.

Whenever a trigger is reached the output signals of  each column and row of the SiPM are summed through the carrier board 
(AiT AB424T) producing 12 row signals and 12 column signals – $Ax_i$ and $Ay_i$. 
There
 are four final outputs $X^+$, $X^-$, $Y^+$, $Y^-$.
 They are functions of  
functions of $Ax_i$ and $Ay_i$ in the following way:

   \comment{
        \begin{equation}
            \begin{gathered}
                X^+ = \sum_i{c_i * Ax_i} \\
                X^- = \sum_i{c_{12 - i} * Ax_i} \\
                Y^+ = \sum_i{c_i * Ay_i} \\
                Y^- = \sum_i{c_{12 - i} * Ay_i}
            \end{gathered}
        \end{equation}
    }

    \begin{equation}
        X^+ = \sum_i{c_i * Ax_i},~~~~~
        X^- = \sum_i{c_{12 - i} * Ax_i}
    \end{equation}

    \begin{equation}
        Y^+ = \sum_i{c_i * Ay_i},~~~~~
        Y^- = \sum_i{c_{12 - i} * Ay_i}
    \end{equation} 

    \noindent such that $c_i + c_{12-i} = 1.0833$, leading to 
    
    \begin{equation}
        {X^+} + {X^-}  = 1.0833 \times \sum_i{Ax_i}
    \end{equation}

    \begin{equation}
        {Y^+} + {Y^-}  = 1.0833 \times \sum_i{Ay_i}
    \end{equation}
    
Each set of output signals is fed into two 16 channel receivers, which act as amplifiers of the signals and also produce analog sum signal.
The count rate of both channel receivers is measured via the quad counter in order to ensure that both of the receivers operate with the same amplification.

Analog sum signals are then passed to a V1742 Digitizer via MCX- MCX cables and recorded if over a set threshold (10 mV).  Total number of samples is set to 1024 with a sampling rate of 1 GSps. 
A DAQ software manages data collection by configuring digitizer polling and recording the data into output files with a fixed format and size.  The analog sum signal produced from each of the boards is connected to a CFD(Constant Fraction Discriminator), 
producing a NIM signal,
which is than fed into the logic unit. 
Upon a coincidence of the input signals another NIM signal is produced and through a MCX-LEMO cable is fed into the digitizer and acts as a local trigger used for coincidence measurement.

\subsection{Front end}
The front end has the same construction as the back end but here the thin crystal is utilized.
The SiPM matrix is connected to a SenSL adapter .
The current signal of the SiPM goes to the SenSL adapter, which routes the signals from the array to header  pins via mating connectors. 
SMA connectors on the board are connected via the supplied jumper cable to any of the array header pins and used for accessing signals or supplying bias voltage.
The second SenSl adapter is planned to be connected to a readout module DT5202 . 
%figure …..
Entering the readout module the signals are amplified by the preamplifier to a  level suitable for further processing, while maintaining the signals integrity.
After amplification the signals are processed by two shapers.
The slow shaper filters and integrates the signal over a longer time to measure the total charge, which correlates to the energy of the input signal.
The fast Shaper  processes the signal quickly, preserving timing information for precise event timing. 

The output of the Slow Shaper is fed to a Peak Stretcher, which captures and holds the peak value of the signal,which represents the total charge (energy).
The output of the fast shaper is fed to the discriminator, 
which compares the signal to the threshold level set by the trigger threshold DAC.
If the signal exceeds the threshold, it generates a trigger to indicate a valid event. 
The trigger signals from all 32 channels can be combined using an OR logic gate to form a global trigger.
The signals from all 32 SiPM channels are sent to a multiplexer,which selects one channel at a time and forwards the signal to the ADC for digitization.
The ADC  converts the analog peak value from the Peak Stretcher into a digital value. 
This digital value represents the energy of the detected signal. 

The FPGA, along with a low-resolution TDC (0.5 ns LSB), is also responsible for processing the discriminator outputs in timing measurements. 
FPGA also collects the digitalized data from both converters and  formats the data for readout, while configuring the ASIC. 
The collected data  is than 
forwarded through ETHERNET to a PC,
in which a  DAQ software handles the data 
collection and its storage into output files with fixed size and format.

\section{Monte Carlo study of the energy deposit and efficiency}

The description of the major components of the experimental setup were
implemented in Geant4 simulation package. 
This includes the front and the back end detection plane.
The simulation  was performed with the gamma line of \({}^{137}\text{Cs}\) (662 keV) for relative distances 5 and 10 cm. 
Given the relative distance, the maximum allowed angle of Compton scattering is therefore $\theta_1 = 26.57^\circ $ for relative distance 5 cm and $\theta_2 = 45^\circ $ 
for relative distance 10 cm. This estimation is made for the  extreme case when gammas fall perpendicularly to the plane of the front end.
Given this information the total energy of the initial photon can be reconstructed using the deposited in the front end crystal, 
where Compton scattering occurs, 
and in the back end crystal,
where the scattered photon is absorbed. 
After calculation, 
at  $\theta_1 = 26.57^\circ $ the maximum deposited energy  in the front end crystal is $E_1 = 79.68$~keV   
and the minimum deposited energy  in the back end  crystal is  $E_2 = 582.33$~keV.
For $\theta_2 = 45^\circ $ it is estimated that the energies are $E_1 = 182.10$~keV and $E_2 = 479.90$~keV.
The Monte Carlo simulation shows that Compton scattering of the initial photon in the first crystal followed by absorption in the second crystal is not the only possible interaction pathway.
\begin{figure}[!htp]
    \begin{center}
        \includegraphics[width=0.47\textwidth]{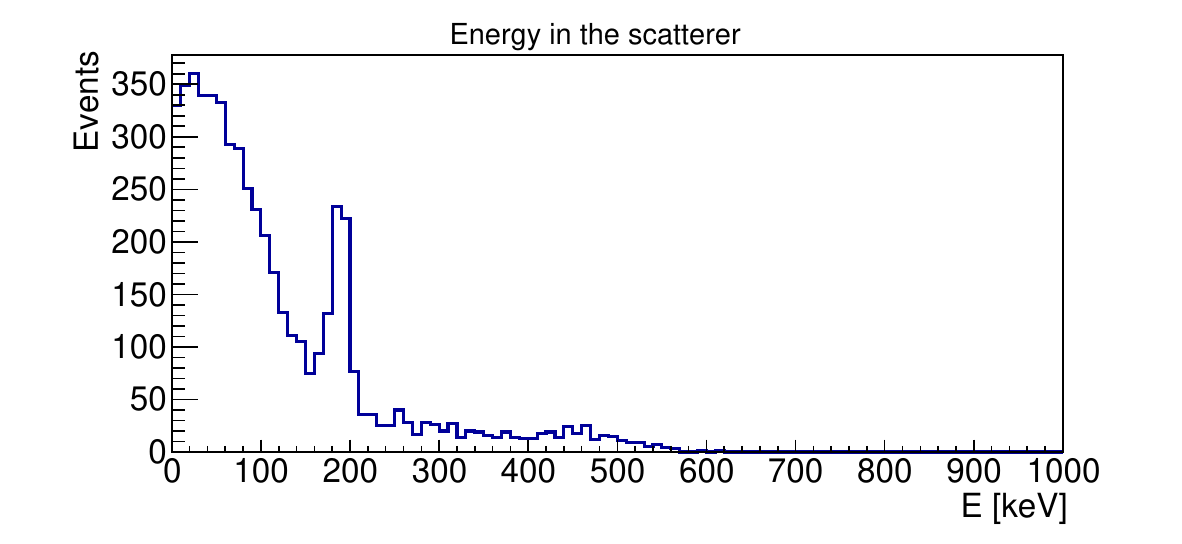}
        \includegraphics[width=0.47\textwidth]{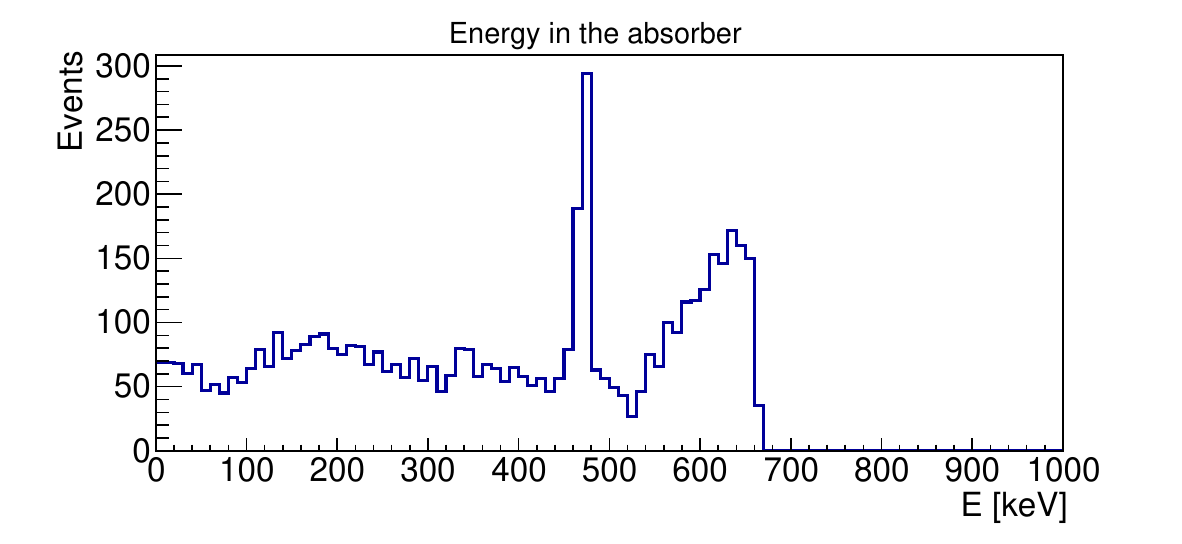}
        \includegraphics[width=0.47\textwidth]{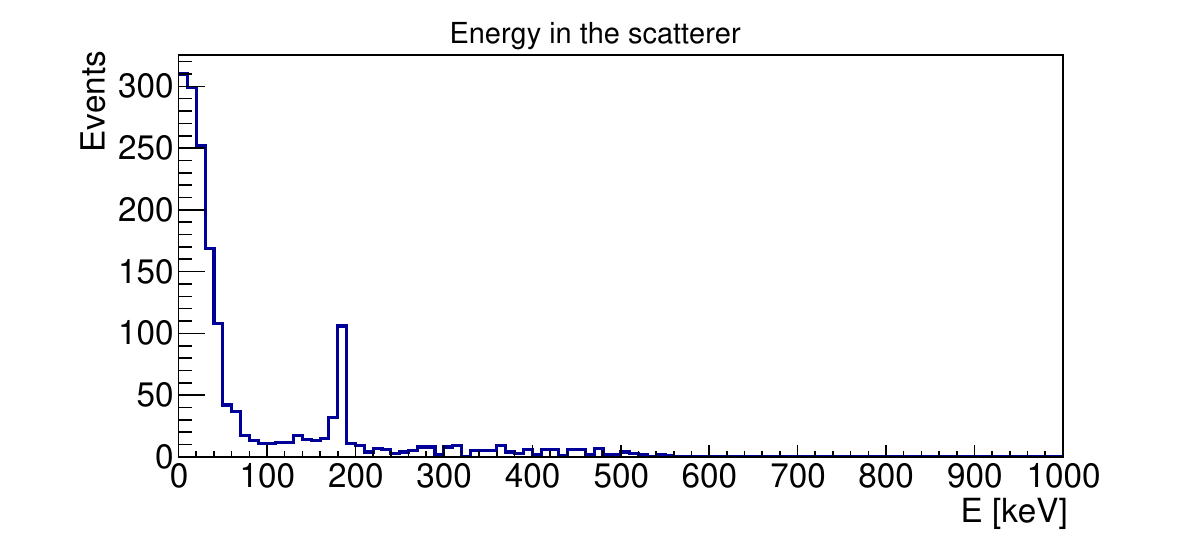}
        \includegraphics[width=0.47\textwidth]{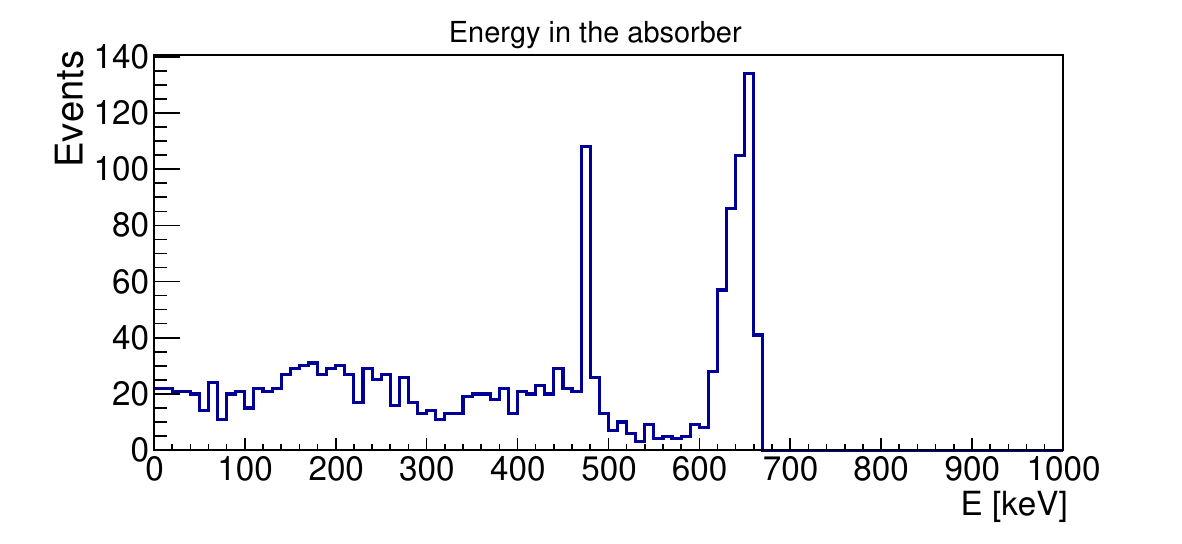}

        \caption{\label{fig:edep} Distribution of the energy deposit in the front (left) and in the back end (right) scintillator for 5 cm (top) and 10 cm (bottom) distance between the scintillators.}
    \end{center}
\end{figure}

\begin{figure}[!htp]
    \begin{center}
        \includegraphics[width=0.45\textwidth]{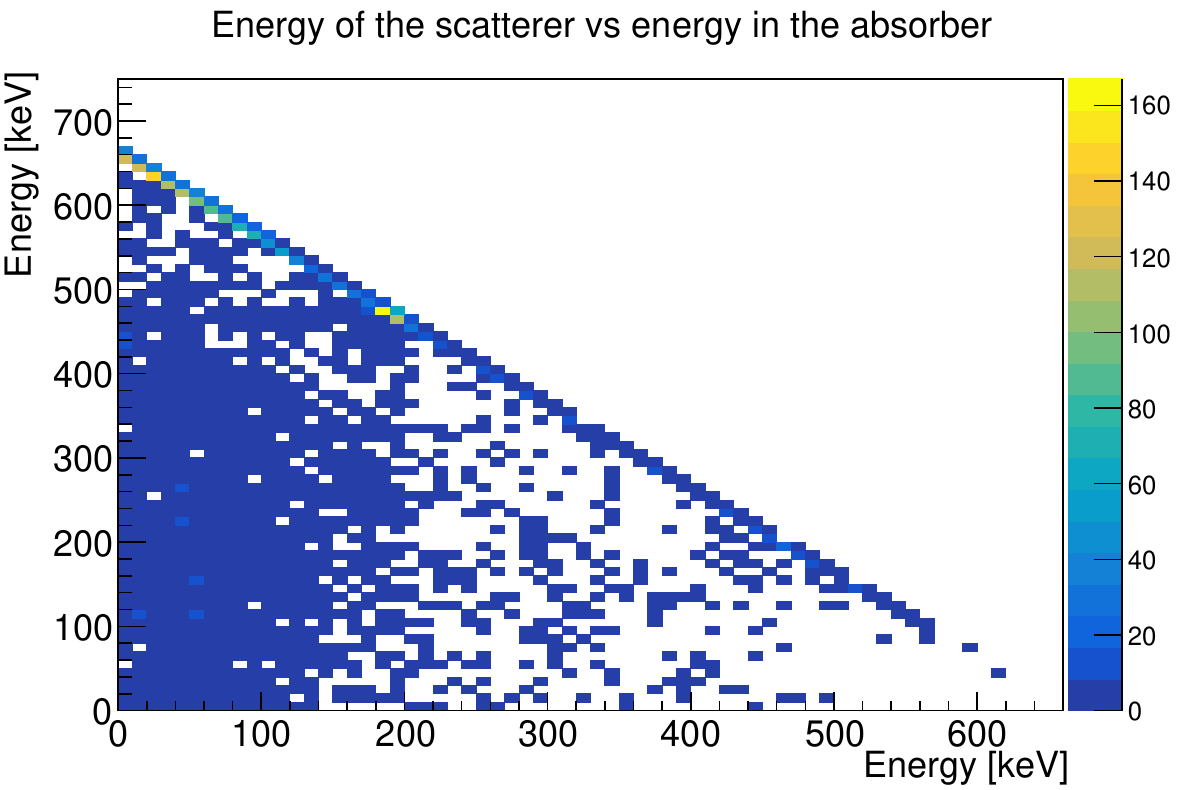}
        \includegraphics[width=0.45\textwidth]{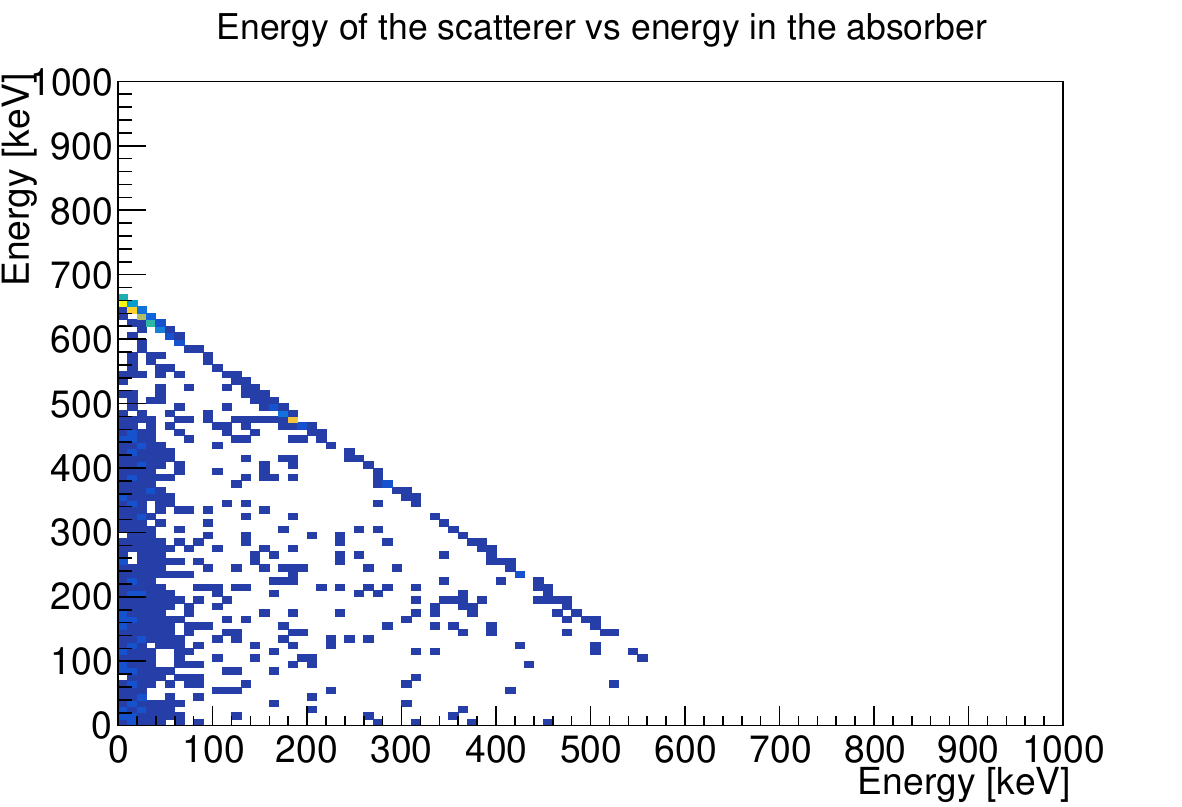}
        \caption{\label{fig:evse} $E_1$ vs $E_2$  for 
        the two relative 
relative distance  5 cm
        (left) and 10cm (right)        
        }
    \end{center}
\end{figure}

The results of the analysis (Figure \ref{fig:edep} ) show that 
there are three types of populations of events in the presented plots:
\begin{itemize}
    \item First type: The initial gamma quant is  Compton scattered in the first crystal (scatterer)  and absorbed in the second one (absorber);

    \item Second type: The initial gamma quant is Compton scattered from the first crystal, but instead of being absorbed in the second crystal it Compton scatters again at angle of 180$^\circ$ and it is absorbed in the first crystal interaction in the second crystal by Compton scattering, 
    back-scattered and absorbed in the first crystal;

    \item Third type: The initial gamma quant is Compton scattered in the first crystal and then Compton scattered in the 
    second crystal,therefore leaving the detector system without being absorbed.
\end{itemize}

The first two populations provide total energy $E = E_1 + E_2$ equal to the 
impinging gamma energy, while the third population 
contributes to a smooth background of events. 

In the case for relative distance 5 cm  in the scatterer the first peak observed corresponds to the events of the population of the first type and the maximum energy of the interaction is at the peak tail's end at $E_{max} = 180  $~keV.
The second peak in the plot corresponds to the events in the second type where the gamma quant undergoes scattering of $180 ^\circ$ and deposits its energy in the other detector plane. 
The energy of this peak is $E_1 = 186.27\pm 0.62~keV $.
This peak  has a better energy resolution than the first peak that is only due to Compton scattering, which allows for better  energy determination of the incident gamma quant.
For the absorber, 
the same conclusions can be drawn except that here the peak attributed to the population of events of second type  has an  energy of $E_2 = 469.79\pm 0.50 ~keV$.
This is followed by the peak corresponding to the population of events of the first type which has minimum limit of the energy at $E_{min} = 479 ~keV$.
In the case of relative distance of 10 cm the  deposited energies from the population of events from the second type are at around $E_1 = 183.46 \pm 0.46 ~keV $ and  $E_2 = 475.34 \pm 0.47~keV $ and  the limited deposited energies for the population of events from the first type  are at around $E_{max} = 80 ~keV$ and $E_{min} = 582~ keV$
%If the total energy of the system is plotted Figure \ref{fig:esum} graphs are compared, it can be seen that for relative distance of 5 cm a much better acceptance is obtained.%

On Figure \ref{fig:evse}   $E_1$ vs $E_2$  for the two relative distances
between the scintillating crystals is presented. 
As it can be seen  a big portion of the energy of the  gamma quant is mainly deposited in the back end where the absorber crystal is, 
while a little fraction is deposited in the front crystal where the scattering occurs.
The plot also shows that there is a higher probability for the front end crystal to receive lower amount of energy than the back end crystal.
For a relative distance of 5 cm many more absorption events are recorded, 5335  versus 1663 events for 10 cm,
out of 100 000 simulated.
The efficiency $\varepsilon$ with which the  Monte Carlo simulated setup determines the initial gamma quant energy can be defined as the ratio of the number of events where the total energy of the initial gamma quant to the total number of event. In that case for a relative distance of 10 cm the efficiency is  $\varepsilon = 0.40 $,
while for a relative distance of 5 cm $\varepsilon = 0.45 $.
The field of view of the detector is represented by the solid angle $\Omega $.The value of the solid angle for relative distance of 5 cm is $\Omega \approx 1~sr $ and for relative distance of 10 cm is  $\Omega \approx 0.25~sr $.
These values along with the values of the efficiency show that at relative distance of 5 cm there is a higher gamma quant interactions with the plane due to the larger angular acceptance.  
%description of the back Compton scattering, 

From the above analysis of the Monte Carlo simulation it can be concluded that events from the second type provide a better energy and time resolution, while  the efficiency with which the detector  performs  energy and thus precise scattering angle $\theta$ reconstruction improves with decreasing relative distance between detection planes.

%
%\begin{figure}[!htp]
%    \begin{center}
%        \includegraphics[width=0.45\textwidth]{4.pdf}
%        \includegraphics[width=0.45\textwidth]{14.pdf}
%        \caption{\label{fig:esum} Sum of the energy in the two scintillators for 5 cm (left) and 10 cm (right) distance. }
%    \end{center}
%\end{figure}

\section{Conclusions}
The simulation shows  the energy 
%and position 
reconstruction capabilities of the experimental setup for a given energy of the gammas.
Out of the three population of  events  the population of the second type are easily distinguishable as a signal and provide a decent way to determine the energy of the initial photon compared to the events of first and third type.
Those type of events are also important for time resolution, because  when the photon undergoes a 180$^\circ$ from one detector plane and it is absorbed in the other, the time it takes for the photon to travel before being absorbed in the plane allows for a better localization of event in time.

The results indicate that a Compton telescope utilizing  interactions with 180$^\circ$ scattering from the second detector can be advantageous for system design as it provides 
better event discrimination capabilities due to the 
fixed ratio of the deposited energy in the two scintilators for
a given energy of the impinging high energy gamma-ray.
%and directionality measurement. 
%, as it enables a more compact geometry, because it enables full energy deposition in a smaller space.
The  test setup was built primarily using modular and commercial
equipment and 3D printed elements. 
It is ready to be exposed to a gamma ray source to validate its applicability 
to operate as a small-scale gamma ray telescope. 

\section*{Acknowledgments}
The authors thank Venelin Kozhuharov 
for continuous support, and for the ideas on the setup and data analysis. 
Sofia University is supported
 by the European Union-NextGenerationEU, through the 
National Recovery and Resilience Plan of the Republic of Bulgaria, 
project SUMMIT BG-RRP-2.004-0008-C01, 
%GB and VA recognize support from the
%Bulgarian Ministry of Education and Science, 
%within the National Roadmap for Research Infrastructures (object CERN),
while SI recognizes support by 
Project BG16RFPR002-1.014-0006 
''National Centre of Excellence Mechatronics and Clean Technologies'', co-funded by the European Union, under ''Research Innovation and Digitization for Smart Transformation'' program 2021-2027.

\end{document}